\documentclass[
    twocolumn,
	prd,
	amssymb,
	preprintnumbers,
	secnumarabic,
	nofootinbib]{revtex4-1}

\pdfoutput=1

\usepackage{graphicx}
\usepackage{enumitem}
\usepackage{latexsym}
\usepackage{amsfonts}
\usepackage{amssymb}
\usepackage{color}
\usepackage{amsmath}
\usepackage{slashed}
\usepackage{dcolumn}
\usepackage{verbatim}
\usepackage{float}
\usepackage{multirow}
\usepackage{xspace}
\usepackage[normalem]{ulem}
\usepackage[
pdfauthor={Nirmal Raj}]{hyperref}

\newcommand{\lsim}{\lesssim}

\def\Oc{\mathcal{O}}

\newcommand{\acro}[1]{\textsc{\MakeLowercase{#1}}} %FOR `NOT SHOUTING' CAPS (e.g. acronyms)
\newcommand{\osn}{\oldstylenums}
 
\newcommand{\beq}{\begin{equation}}
\newcommand{\eeq}{\end{equation}}
\newcommand{\bea}{\begin{eqnarray}}
\newcommand{\eea}{\end{eqnarray}}
\newcommand{\nn}{\nonumber}

\def\GF{G_{\rm F}}

\def\ER{E_{\rm R}}
\def\mT{m_{\rm T}}

\def\dstar{d_*}

\definecolor{orange}{rgb}{1,0.5,0}
\definecolor{purple}{rgb}{1,0,1}
\definecolor{brown}{rgb}{.7,.2,.2}
\definecolor{violet}{rgb}{.6,.3,.8}
\definecolor{nicegreen}{rgb}{.3,.7,.3}

\allowdisplaybreaks

\begin{document}

\title{Neutrinos from Type Ia and failed core-collapse supernovae at dark matter detectors}

\author{Nirmal Raj}
\affiliation{TRIUMF, 4004 Wesbrook Mall, Vancouver, BC V6T 2A3, Canada}

%%%%%%%%%
\begin{abstract}
Neutrinos produced in the hot and dense interior of the next galactic supernova would be visible at dark matter experiments in coherent elastic nuclear recoils.
While studies on this channel have focused on successful core-collapse supernovae, a thermonuclear (Type Ia) explosion, or a core-collapse that fails to explode and forms a black hole, are as likely to occur as the next galactic supernova event.
I show that generation-3 noble liquid-based dark matter experiments such as \acro{darwin} and \acro{argo}, 
operating at sub-keV thresholds with ionization-only signals, would distinguish between 
(a) leading hypotheses of Type Ia explosion mechanisms by detecting an $\Oc$(1)~s burst of $\Oc$(1) MeV neutrinos, and
(b) progenitor models of failed supernovae by detecting an $\Oc$(1)~s  burst of $\Oc$(10) MeV neutrinos, especially by marking the instant of black hole formation from abrupt stoppage of neutrino detection.
This detection is sensitive to all neutrino flavors and insensitive to neutrino oscillations, thereby making measurements complementary to neutrino experiments.

\end{abstract}
%%%%%%%%

\maketitle

%%%%%%%%%%%%%%
\begin{figure*}
\includegraphics[width=.45\textwidth]{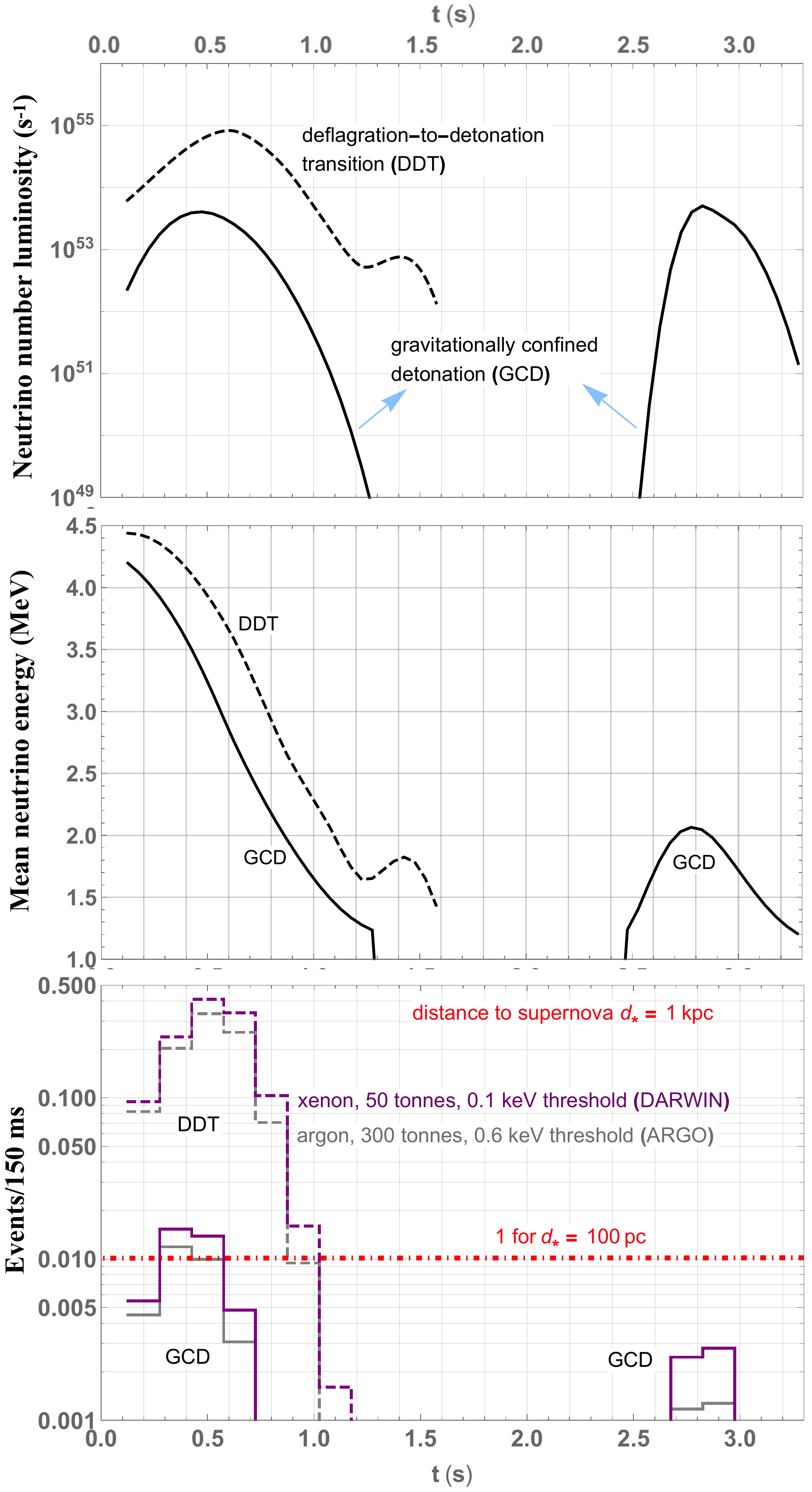} \quad
\includegraphics[width=.49\textwidth]{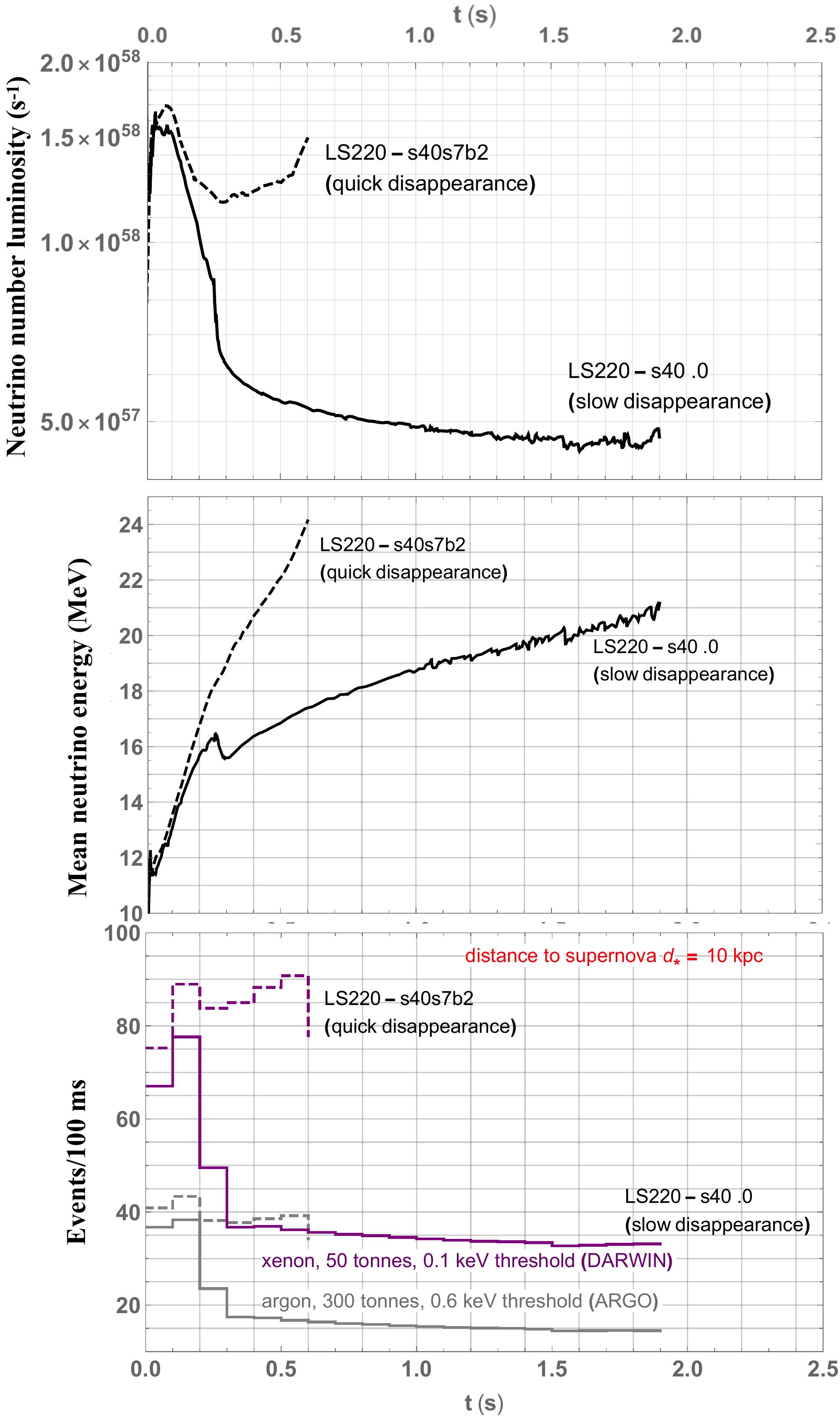} 
\caption{
\textbf{{\em Left}}: Type Ia supernovae, 
\textbf{{\em right}}: failed core-collapse supernovae. 
Shown as a function of time are 
number luminosities of supernova neutrinos (summed over all flavors) emitted at the source ({\bf \em top}), 
mean neutrino energies for all flavors combined ({\bf \em middle}), and 
events per binned time at generation-3 dark matter detectors ({\bf \em bottom}).
Explosion mechanisms of Type Ia supernovae and progenitor models of failed supernovae are visibly distinguished by these detectors.
The wiggles in the top-right and middle-right plots reflect those in Ref.~\cite{1508.00785}.
} 
\label{fig:rates}
\end{figure*}
%%%%%%%%%%%%%%%

%%%%%%%%%%%%%%
\begin{figure*}
\includegraphics[width=.478\textwidth]{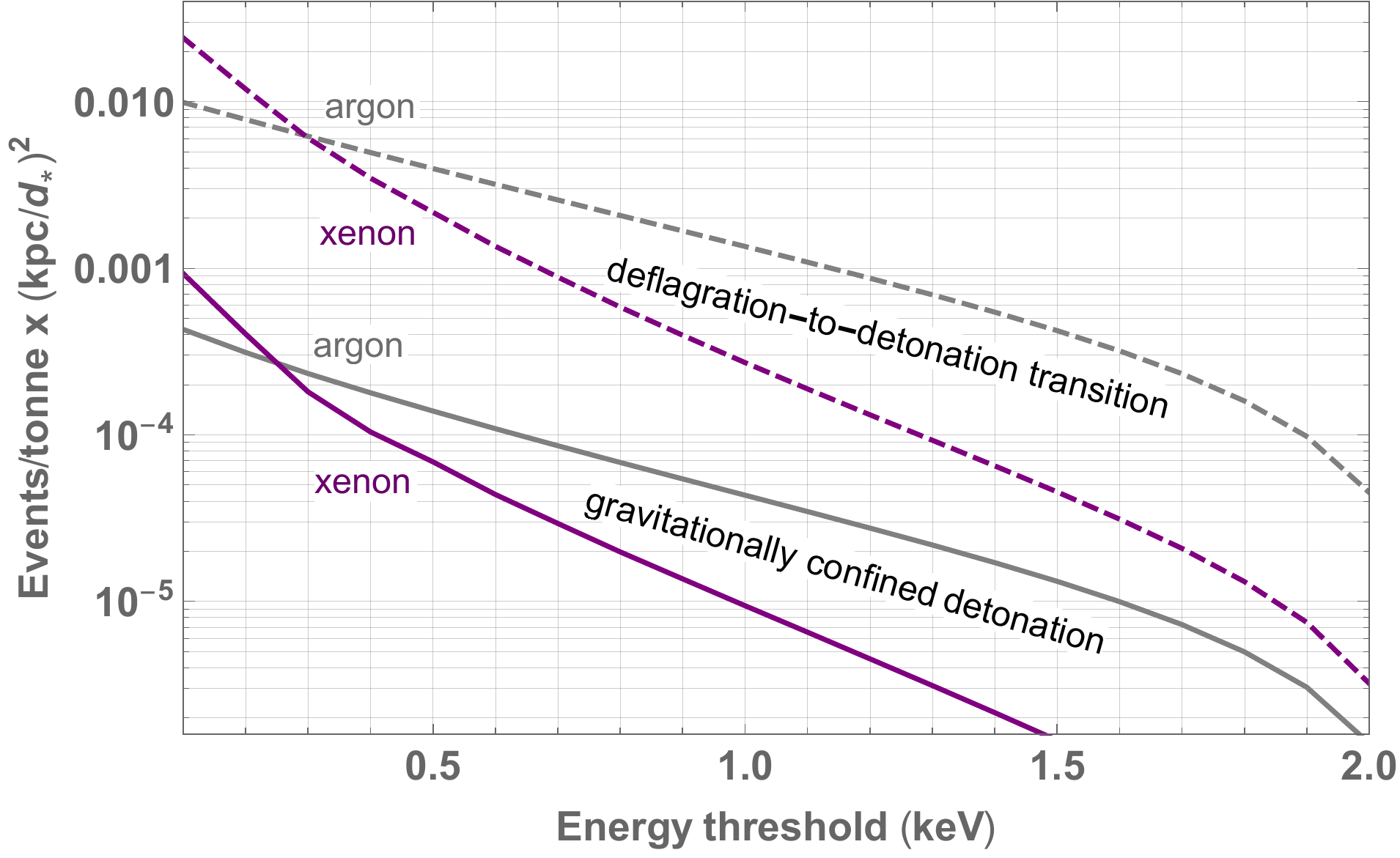} \quad
\includegraphics[width=.472\textwidth]{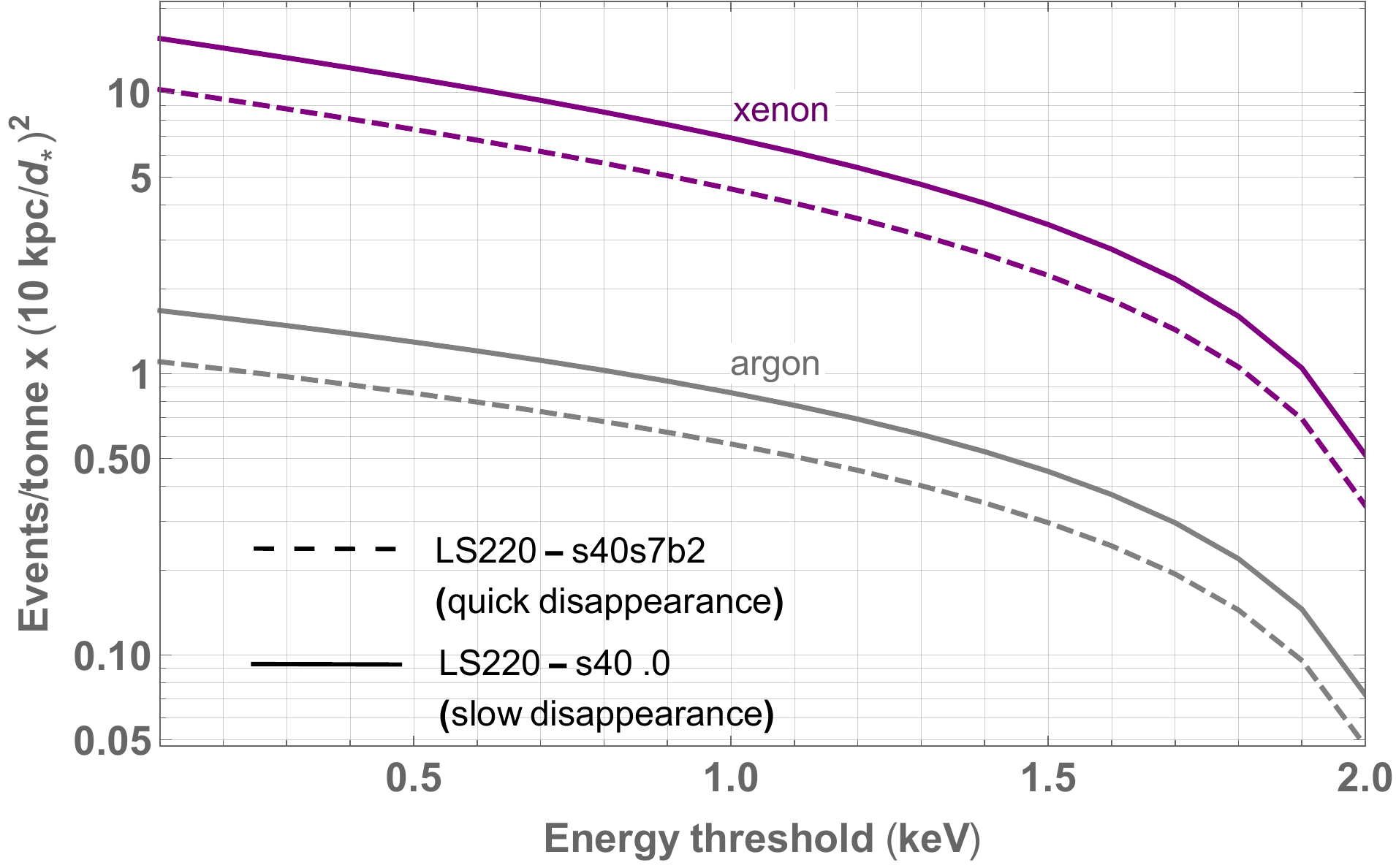}
\caption{
Events detected per tonne of target as a function of detector energy threshold for neutrinos from Type Ia (\textbf{{\em left}}) and  
failed core-collapse (\textbf{{\em right}}) supernovae,
normalized to a supernova distance of 1 kpc and 10 kpc respectively.
}
\label{fig:threshold}
\end{figure*}
%%%%%%%%%%%%%%%

The next galactic supernova is imminent.
This could be induced by thermonuclear runaway fusion (Type Ia supernova) or a rapid core-collapse, estimated to occur at a rate of respectively 1.4$^{+1.4}_{-0.8}$ and 3.2$^{+7.3}_{-2.6}$ per century~\cite{1306.0559}.
The core-collapse often successfully blows away accreting outer layers and leaves behind a neutron star, such as believed to have happened in the last observed galactic supernova \acro{SN} \osn{1987}\acro{A}; yet 10\%-50\% of them fail to explode, unable to prevent intense accretion, leaving behind a black hole\footnote{Violent explosions collapsing into themselves also leave behind other singularities~\cite{mazama}.}.
In all types of supernovae, their hot and dense environments produce neutrinos that escape them and serve as the first particle messengers of this once-in-a-lifetime event.
The neutrino signal would inform whether, when, and where to look for the electromagnetic signal,
and would reveal vital information about the explosive conditions of the progenitor, of which there is currently little measurement or consensus.
Neutrino experiments such as 
IceCube, 
Hyper-K, 
\acro{DUNE}, 
\acro{JUNO}, 
and \acro{HALO} 
will be prepared to detect supernova neutrinos in a range of channels~\cite{1205.6003,1508.00785}, but lately it has been recognized that dark matter experiments, designed for detecting coherent elastic nuclear recoils, are an equally important player capable of uncovering complementary physics.
Whilst studies have been performed on elastic nuclear recoils produced by neutrinos from successful core-collapse supernovae~\cite{
Beacom:2002hs,
Horowitz:2003cz,
1103.2768,
1309.4492,
1604.01218,
1606.09243,
1801.05651,
1806.01417,
1906.05800} 
and pre-supernova nuclear burning~\cite{1905.09283}, 
they are lacking for neutrinos from Type Ia and failed core-collapse.
The purpose of this note is to close these gaps, and to comment on this  detection channel vis-\`a-vis those at neutrino experiments.

Neutrinos and dark matter experiments are intimately connected.
The ``direct detection" program began when a proposal to detect neutrinos via coherent elastic scattering~\cite{Drukier:1983gj} --
 a process observed only recently~\cite{1708.01294,1907.12444} -- 
 was adapted for dark matter searches~\cite{Goodman:1984dc}.
With ever-increasing exposure, these experiments would eventually run into an irreducible background from solar, atmospheric, and relic supernova neutrinos, the ``neutrino floor".
These experiments could also dedicate searches to neutrinos from various sources (including dark matter)~\cite{
1409.0050,
1506.08309,
1510.04196,
1604.01025,
1712.06522,
1801.10159,
1807.07169,
1501.03166,
1711.04531,
1812.05102,
1812.05550}.
While neutrino experiments -- whose detection is usually restricted to only the flavors $\nu_e$ and $\overline{\nu}_e$  -- have larger exposures, dark matter experiments could compensate with enhanced detection rates due to nuclear coherence, and by detecting all flavors ($\nu_e, \overline{\nu}_e, \nu_\mu, \overline{\nu}_\mu, \nu_\tau, \overline{\nu}_\tau$).
This latter feature enables them to reconstruct a supernova neutrino burst without uncertainties from neutrino oscillations in the supernova, and to measure the energy emitted in each flavor.

The supernova neutrino phase space is best sampled by detectors that are large and operating at low thresholds. 
I will thus compute event rates at future generation-3 detectors\footnote{Should a galactic supernova occur during the running of current or next-generation dark matter experiments, my event rates may be trivially rescaled by the target mass.}: 
the xenon-based \acro{darwin}~\cite{1606.07001} and
argon-based \acro{argo}~\cite{1707.08145}. 
Projected with $\Oc$(100)-tonne target mass, these are said to be ``ultimate" detectors that could probe down to the lowest reachable dark matter-nucleon cross sections and the highest reachable dark matter masses~\cite{1803.08044}. 
These detectors are also capable of very low, sub-keV thresholds, as I will discuss later.
\\

%%%%%%%%%%%%%%%%
\noindent
{\bf \em Explosion characteristics and neutrino fluxes.}
% Now I discuss the supernova progenitors and the consequent neutrino fluxes which % will determine the rates of detection, and compare it with HyperK.
\noindent
\underline{Type Ia supernovae.}
Despite their well-known utility as standard candles that suggest that the universe is accelerating \cite{Riess:1998cb,Schmidt:1998ys,Perlmutter:1998np}, little is known about how Type Ia supernova progenitors explode \cite{1302.6420}, or even what they are, although it is argued that they are carbon-oxygen white dwarfs accreting mass from a binary companion that triggers explosive carbon burning.
Determining the explosion mechanism from extragalactic supernovae will be challenging due to telescope limitations, but a supernova in the Milky Way would help settle the question via not only electromagnetic signals, but also neutrinos and gravitational waves.
In particular, neutrinos -- produced through $e^+e^-$ annihilations and $e^{\pm}$ capture on nucleons and nuclei, and carrying away $\sim$1\% of the star's gravitational binding energy~\cite{1511.02542} -- could distinguish between explosion mechanisms even if the electromagnetic signals are alike.
Since the Type Ia supernova core is not dense enough to trap neutrinos, their flux is reliably computed as there is no neutrino transport or self-interactions to account for, unlike for core-collapse supernovae.
  
References~\cite{1605.01408,1609.07403} computed these fluxes using 3D simulations of near-Chandrasekhar mass white dwarfs for two leading hypotheses of the explosion mechanism.
In the first mechanism, deflagration-to-detonation transition (\acro{ddt}), the flame front of subsonic combustion of the fuel -- deflagration -- reaches low density regions, whereupon turbulence shreds it, and cold fuel and hot ashes mix.
This triggers supersonic combustion -- detonation -- of the remaining fuel.
In the second mechanism, gravitationally confined detonation (\acro{gcd}), 
deflagration ash floats to the star surface, but is kept from escaping by the star's gravity, whereupon it envelopes the surface, converges, compresses, and detonates the rest of the fuel.
As neutrinos propagate through the supernova medium they oscillate, and their flavor composition at emission would depend on both the density profile along the line of sight (as the explosion is asymmetric), and on neutrino mass ordering. 
However, for detection via elastic nuclear scattering flavors are not relevant, only the total flux is.
I use tables of neutrino fluences provided by J.~Kneller~\cite{kneller} 
(parts of which are plotted in \cite{1605.01408,1609.07403}) 
to compute energy-differential number luminosities (in units of s$^{-1}$MeV$^{-1}$) summed over all flavors, as a function of post-explosion time.
Dividing by $4\pi d^2_*$, where $d_*$ is the distance to the supernova, gives the evolving differential flux (in cm$^{-2}$s$^{-1}$MeV$^{-1}$) received on Earth, $d^2\Phi/dE_\nu dt$.
In the top left panel of Fig.~\ref{fig:rates} I show the energy-integrated number luminosity versus time for the two explosion mechanisms.
The small second peak in the \acro{DDT} luminosity arises from $e^-$ capture on copper; the two distinct peaks in the \acro{gcd} luminosity correspond to neutrinos produced during deflagration and detonation of the fuel, the valley between them caused by there being no regions hot enough to be in nuclear statistical equilibrium~\cite{1609.07403}. 
In the middle-left panel I plot the mean neutrino energy, which shows that the spectrum generally softens with time.

White dwarf binary mergers (``collisional double-degenerate progenitors") may also cause Type Ia supernovae~\cite{1111.4492,1312.0628,1403.4087}, which could be probed by gravitational wave signals~\cite{1511.02542}. 
To the best of my knowledge, the associated neutrino fluxes have yet to be computed from simulations;
the neutrino signal could help distinguish this scenario from the one considered above, an interesting exercise that I reserve for future study.

\noindent
\underline{Failed supernovae.} 
Neutrinos from core-collapses that fail to explode could constitute 50\% of the relic supernova flux~\cite{0901.0568,1012.1274}, 
may have helped select amino acid chirality~\cite{Boyd:2010ak},
and would increase the amount of technetium-97 in molybdenum ores~\cite{0901.0581}.
As these core-collapses form black holes within $\sim$1 s, they cannot be picked up by telescopes, however, following the suggestion of Ref.~\cite{0802.0456} to monitor supergiants, a star each was seen to disappear in 
real time~\cite{1411.1761} and
archival data~\cite{1507.05823}.  
While these extragalactic observations provided useful constraints, a failed supernova in the galaxy would further offer a wealth of science in the form of neutrinos.

Neutrinos are produced in core-collapse supernovae from $e^+e^-$ annihilations and neutronization, and carry away 99\% of the star's gravitational binding energy. Whereas neutrinos from successful core-collapses diffuse out of the proto-neutron star over $\Oc$(10)~s (the duration of the neutrino signal detected), those from failed supernovae diffuse over $\Oc(1)$~s with a progressively hardening spectrum before the emission abruptly stops due to black hole formation.
These neutrinos are overall harder and brighter than in successful supernovae due to the increase in temperature and density from accretion of matter.
Exactly when the black hole forms, and how the spectrum evolves, depend on progenitor properties like stellar mass and distributions of density, temperature and electron fraction~\cite{0808.0384}, 
as well as on the equation of state of matter at nuclear densities~\cite{0706.3762}. 
For this study I will compare two 40$M_\odot$ progenitor models with the \acro{ls}\osn{220}~\cite{Lattimer:1991nc} equation of state: 
s\osn{40}s\osn{7}b\osn{2}~\cite{Woosley:1995ip}, disappearing ``quickly" in 0.6~sec, and 
s\osn{40}.\osn{0}~\cite{Woosley:2002zz}, ``slowly" in 1.9~sec.
The differential flux for flavor $\alpha$ at supernova distance $d_*$ is
%%%%%%%%
\beq
 \frac{d^2\Phi_\alpha}{dE_{\nu_\alpha}dt} =
 \frac{1}{4\pi \dstar^2}
 \frac{L_{\nu_\alpha}(t)}{\langle E_{\nu_\alpha}(t) \rangle} 
\varphi_\alpha(E_{\nu_\alpha},t),
\eeq
%%%%%%%%
where
$L_{\nu_\alpha}$ and $\langle E_{\nu_\alpha} \rangle$ are the luminosity and mean energy, whose values I take from Ref.~\cite{1508.00785},
and 
$\varphi$ is the normalized energy spectrum parameterized by~\cite{Keil:2002in}
%%%%%%%%%
\bea
\varphi_\alpha(E_{\nu_\alpha},t) &=& 
\nn \langle E_{\nu_\alpha}(t) \rangle^{-1}
\frac{(1+\xi_\alpha)^{1+\xi_\alpha}}{\Gamma(1+\xi_\alpha)}
 \left(\frac{E_{\nu_\alpha}}{\langle E_{\nu_\alpha}(t) \rangle}\right)^{\xi_\alpha}   \times \\
& & \exp\left(-\frac{(1+\xi_\alpha)E_{\nu_\alpha}}{\langle E_{\nu_\alpha}(t) \rangle}\right)~,
\eea
%%%%%%%%%
where $\xi_\alpha \equiv \xi_\alpha(t)$ is related to the ``pinching parameter" $p$ by  $p = 1.303^{-1}(2+\xi)/(1+\xi)$. 
The $p(t)$ for \acro{ls}\osn{220}-s\osn{40}.\osn{0} are taken from Fig.~3.15 of Ref.~\cite{garchthes}, and these are assumed the same for \acro{ls}\osn{220}-s\osn{40}s\osn{7}b\osn{2}. 
In the top and middle right panels of Fig.~\ref{fig:rates} I show the evolution of the number luminosity and mean energy (combined for all flavors). 
The quickly-disappearing supernova neutrinos are brighter and harder, which as we will see, would result in higher detection rates.
\\

%%%%%%%%%%%%%%%%
\noindent
{\bf \em Prospects at dark matter detectors.}

With the neutrino fluxes in hand, I now compute the differential scattering rate (per tonne of detector mass) as  
%%%%%%%%%%%%%%%%
\beq
\frac{d^2R}{d\ER dt} =
N^{\rm ton}_{\rm T} 
\int_{E_\nu^{\rm min}} 
dE_\nu \frac{d^2\Phi}{dE_\nu dt} \frac{d\sigma}{d \ER}~,
\label{eq:ratescat}
\eeq
%%%%%%%%%%%%%%%%
where $N^{\rm ton}_{\rm T}$ = 4.57$\times$10$^{27}$ (1.51$\times$10$^{28}$) is the number of nuclei per tonne of liquid Xe (Ar), and
$E_\nu^{\rm min} = \sqrt{\mT\ER/2}$ is the minimum $E_\nu$ required to induce a nuclear recoil of energy $\ER$.
The differential scattering cross section for a nuclear target\footnote{Scattering on electrons will be highly subdominant~\cite{1604.01218}.} with 
mass $\mT$, 
$N$ neutrons and 
$Z$ protons is given by \cite{Freedman:1977xn}
%%%%%%%%%%
\bea
\nn \frac{d\sigma}{d \ER}(E_\nu, \ER) &=&
\frac{\GF^2}{4\pi} \mT 
[N - (1-4\sin^2\theta_{\rm W})Z]^2 \\
 & &\left(1 - \frac{\mT \ER}{2E^2_\nu} \right)
F^2(\ER)~,
\label{eq:xsscat}
\eea
%%%%%%%%%
where 
$\GF = 1.1664 \times 10^{-5}$ GeV$^{-2}$ is the Fermi constant,
$\sin^2\theta_{\rm W} = 0.2387$ the Weinberg angle at this energy scale, 
and $F(\ER) \simeq 1$ the Helm nuclear form factor~\cite{Helm:1956zz}.
From the fact that the kinematics-limited maximum $\ER^{\rm max}$ = $2E^2_\nu/(\mT + 2E_\nu)$, the total cross section $\propto E_\nu^2$, as expected for
scattering that proceeds via a dimension-6 operator (from integrating out a $Z$ boson mediator). 

To obtain event counts I assume a detector mass of 50 (300) tonnes for \acro{darwin} (\acro{argo}). 
Since the neutrino burst is brief, fiducialization is unnecessary. 
I then integrate Eq.~\eqref{eq:ratescat} over binned time intervals, and from $\ER$ = detection threshold up to 
$\ER^{\rm max}$.
For the threshold I assume 0.1~keV$_{\rm NR}$ (0.6~keV$_{\rm NR}$) for \acro{darwin} (\acro{argo}).
These are realistic possibilities if ionization-only signals are deployed. 
Xenon experiments have by that means achieved 0.7~keV$_{\rm NR}$ threshold~\cite{1605.06262}, and with reduction in backgrounds and sensitivity to single-/double-electron channels anticipated\footnote{I thank Rafael Lang for this intelligence.}, this threshold is expected to lower significantly in generation-3 detectors; 
argon experiments have achieved 0.6~keV$_{\rm NR}$ threshold~\cite{1802.06994} and are expected to re-achieve it in their 3rd generation.
In any case, lest my assumptions turn out overoptimistic, I plot for the reader's reference the net events per tonne of target versus threshold in Fig.~\ref{fig:threshold}, for a Type Ia (failed) supernova at $d_*$ = 1 kpc (10 kpc).

In the bottom panels of Fig.~\ref{fig:rates} I plot events per 150 ms (100 ms) for Type Ia (failed) neutrinos from a distance of 1 kpc (10 kpc).
Due to its higher fluxes and energies,
the \acro{ddt} Type Ia mechanism results in $\sim$50$\times$ more events than \acro{gcd}, clearly separating them.
Due to its softness, the \acro{gcd} detonation peak becomes impossible to detect unless $d_* \lsim 70$ pc, which is unlikely.
The \acro{ddt} event rates are comparable to Super-K, \acro{DUNE} and \acro{JUNO}, and the \acro{gcd} rates are $\sim$ 10$\times$ smaller~\cite{1605.01408,1609.07403}. 
Again due to higher fluxes and energies, the \acro{ls}\osn{220}-s\osn{40}s\osn{7}b\osn{2} failed supernova yields more events than the \acro{ls}\osn{220}-s\osn{40}.\osn{0}.
The clearer distinguisher is the time at which neutrino detection abruptly stops, signifying black hole formation.
In these energy ranges the relevant irreducible background is solar $^8$B neutrinos, but it is negligible at a rate of $\simeq$ 2$\times 10^{-3}$ events/s for my detector configurations~\cite{0903.3630}.
The detector backgrounds are less understood, and estimated to be $\simeq$ 1 event/s for \acro{darwin}~\cite{1606.09243}.
Pileup -- the smearing of arrival times of electrons drifted into the gas phase of the \acro{tpc} -- could limit the timing resolution of events for a sufficiently close supernova.
This will not be an issue so long as events are separated by $\Oc$(ms)~\cite{1606.09243,1707.08145}.
For an even closer supernova, the pulse width of the ionization signal, $\Oc(\mu$s), determines whether individual events may be resolved.
\\

%%%%%%%%%%%%%%%
\noindent
{\bf \em Summary and outlook.}

In this note I have sketched the detection prospects of neutrinos from an imminent galactic Type Ia and failed core-collapse supernova at generation-3 dark matter experiments.
Unlike the optical signal, this detection could distinguish between leading hypotheses of Type Ia explosion mechanisms, deflagration-to-detonation transition and gravitationally confined detonation.
This detection could also identify the progenitor of a failed supernova, in particular clearly marking the time at which the proto-neutron star disappears into a black hole.
Though lacking in the ability to reconstruct neutrino direction and localize the supernova, dark matter experiments would complement neutrino telescopes that typically detect $\nu_e$ and $\overline{\nu}_e$ flavors:
as coherent elastic nuclear scattering is flavor-blind, it is insensitive to neutrino oscillations in the stellar medium and free space, and it could measure the energy distribution across neutrino flavors.
Finally, that the background rates are low for all types of supernova neutrinos furthers the case for adding dark matter experiments to the Supernova Early Warning System~\cite{Antonioli:2004zb,0803.0531}.
\\

\noindent 
Many thanks to James Kneller for e-mailing tables of neutrino fluences from Type Ia supernovae.
I also thank 
Michela Lai,
Rafael Lang,
David Morrissey,
Daniel Siegel, 
and
Shawn Westerdale
for helpful conversations.
This work is supported by the Natural
Sciences and Engineering Research Council of Canada (\acro{nserc}).
T\acro{RIUMF} receives federal funding
via a contribution agreement with the National Research Council Canada.
This work was performed in part at the Aspen Center for Physics, which is supported by National Science Foundation grant PHY-1607611.
%%%%%%%%%%%%%%%%%%%%%%%%

\appendix

\end{document}